\documentclass[10pt, twoside]{article}

\usepackage[utf8]{inputenc}

\usepackage{geometry}
\usepackage{hyperref}
\usepackage{graphicx}
\usepackage{booktabs}
\usepackage{array}
\usepackage{paralist}
\usepackage{verbatim}
\usepackage{subfig}
\usepackage{amsmath}
\usepackage{mathtools}
\usepackage{amsthm}
\usepackage{amssymb}
\usepackage{longtable}
\usepackage{undertilde}
\usepackage{xcolor}
\usepackage{float}
\usepackage{fancyvrb}
\usepackage{multirow}
\usepackage{etoolbox}
\let\bbordermatrix\bordermatrix
\patchcmd{\bbordermatrix}{8.75}{4.75}{}{}
\patchcmd{\bbordermatrix}{\left(}{\left[}{}{}
\patchcmd{\bbordermatrix}{\right)}{\right]}{}{}

\newtheorem{theorem}{Theorem}

\theoremstyle{definition}

\theoremstyle{remark}

\theoremstyle{plain}

\usepackage{fancyhdr}

\usepackage{sectsty}
\allsectionsfont{\sffamily\mdseries\upshape}

\usepackage[nottoc,notlof,notlot]{tocbibind}
\usepackage[titles,subfigure]{tocloft}

\newcommand{\bx}{\boldsymbol{x}}

\newcommand{\btheta}{\boldsymbol{\theta}}
\newcommand{\bphi}{\boldsymbol{\phi}}
\newcommand{\bgamma}{\boldsymbol{\gamma}}
\newcommand{\blambda}{\boldsymbol{\lambda}}
\newcommand{\bzero}{\boldsymbol{0}}

\def\hang{\hangindent\parindent}
\def\rf{\par\noindent\hang}

\newcommand{\bs}{\boldsymbol}

\graphicspath{ {images/} }

\begin{document}

{\large

\newpage
\thispagestyle{plain}
\setcounter{page}{1}

 \section*{\textbf{The large sample coverage probability of confidence intervals \newline in general regression models after a preliminary hypothesis test}}
 
 \medskip

PAUL KABAILA

\noindent \textit{Department of Mathematics and Statistics, La Trobe University Melbourne, Australia}

\medskip

\noindent RUPERT E. H. KUVEKE

\noindent \textit{Department of Mathematics and Statistics, La Trobe University Melbourne, Australia}

\vspace{4cm}

\noindent Running headline: Post-model-selection interval coverage 

\newpage
\thispagestyle{plain}

\noindent
\textbf{ABSTRACT. \
We derive a computationally convenient formula for the large sample coverage 
probability of a confidence interval for a scalar parameter of interest following
a preliminary hypothesis test that a 
specified 
vector parameter takes a given value
in a general
regression model. 
Previously, this large sample coverage probability 
could only be estimated by simulation. 
Our 
 formula only
requires the evaluation, by numerical integration, of either a double or triple integral, irrespective of 
the dimension of this 
specified vector parameter.
We illustrate the application of this formula to a confidence interval 
for the log odds ratio of myocardial infarction when the exposure is
 recent oral contraceptive use, following a preliminary test that two specified interactions in a logistic regression model are zero.
 For this real-life data, we compare
this large sample coverage  probability with the actual 
coverage probability of this 
confidence interval, obtained by simulation.
}
\bigskip
\bigskip

\noindent  \textit{Key words:}
bootstrap, confidence interval, coverage probability, generalized linear models, large sample coverage probability, model selection, post-model-selection confidence interval 

\newpage

\noindent {\bf{1. Introduction}}

\medskip

\noindent 
Preliminary data-based model selection is widespread in applied statistics. Commonly, a preliminary hypothesis test is carried out, followed by the 
construction of a confidence interval for the parameter of interest based on the assumption that the selected model had been given to us \textit{a priori}, as the true model. 
For a linear regression with independent and identically normally distributed errors, there is an extensive literature on the coverage
properties of such a post-model-selection confidence interval. 
For a review of this literature see e.g. Kabaila (2009).
Post-model-selection confidence intervals are still in common use in the context of generalized linear models, see e.g. 
Kabat et al. (2010), O'Donnell et al. (2010), Stampf et al. (2010), Li et al. (2011), Huber et al. (2014), Bendas et al. (2015) and Kanbayashi et al. (2017).
It is therefore important to also assess the coverage properties of the post-model-selection confidence interval in the context of general regression models.
Currently, the most important contribution to this assessment is the expression for the large sample coverage probability given directly below Figure 1  
of Hjort \& Claeskens (2003).

In the present paper we suppose that a preliminary hypothesis test is used to select one of two nested general regression models: 
the full model and a restricted model in which a $q$-dimensional vector parameter takes a specified value.
Of course, for the appropriate test size, this is equivalent to choosing the model that minimizes AIC.
 For $q=1$,  Hjort \& Claeskens (2003) show that 
their expression for the large sample coverage probability is equal to the sum of two one-dimensional integrals, which can be readily evaluated using numerical integration.
However, without further work, the only method available for the evaluation of this expression for $q > 1$ is simulation.
 
Our main result is to show that this expression is equal to a formula consisting of a trivial term added to either a double integral for $q = 2$ or a triple integral for all $q > 2$ 
(Theorem 1).
These multiple integrals, which are readily evaluated by numerical integration, are derived using the methods in the appendix of
Kabaila \& Farchione (2012). 
This formula also possesses a symmetry property 
(Theorem 2)
which halves the time needed to compute the minimum coverage probability.

Throughout the paper, we will refer to the following \textbf{case control example}.
The data for this example is given in Table 7.6 of Schlesselman (1982)
and the parameter of interest is the odds ratio of myocardial infarction (MI) 
in relation to recent oral contraceptive (OC)  use.
For this example,
Schlesselman (1982, p.255) conducts a preliminary test, with large sample size 0.05, of the null hypothesis that the coefficients of two specified second order interaction terms are both zero (i.e. $q = 2$) against the alternative hypothesis that at least one of these coefficients is non-zero.
He accepts this null hypothesis and then constructs the confidence interval  
[1.9699, 5.4799] 
for the odds ratio. 
This confidence interval has nominal coverage 0.95.
To swiftly assess the actual minimum coverage probability of this  
post-model-selection confidence interval, we evaluate its 
large sample coverage probability using 
Theorem 1.
Figure 1 is a contour plot of this large sample coverage probability. 
All of the computations for this paper were carried out using programs written in {\tt R}. 
The minimum large sample coverage probability is $0.3281$, which is far below the nominal coverage, indicating that
Schlesselman's (1982) post-model-selection confidence interval should not be used. 
Instead, the confidence interval, with the same nominal coverage, based on the full model should be used. 
This confidence interval is 
[1.1526, 11.1251],
which covers a substantially wider set of values of the odds ratio.
 
 \medskip

 \begin{figure}[H]	 	
	\includegraphics[width = .95\textwidth]{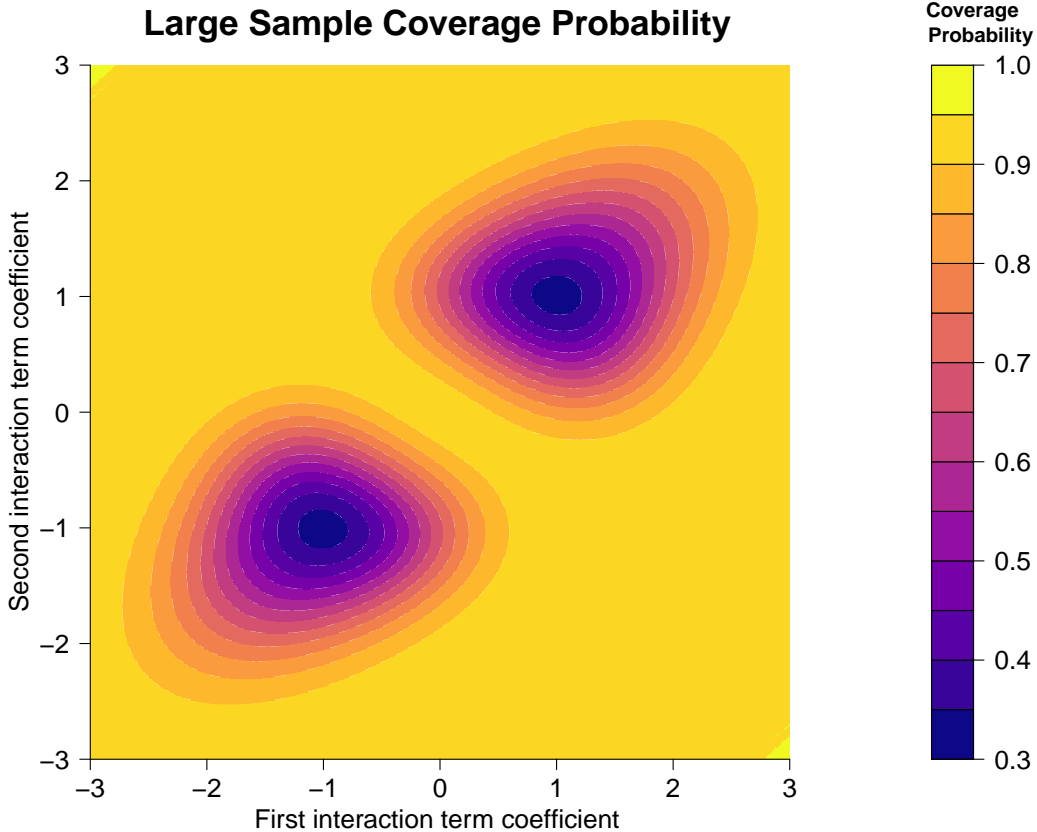}
 	\centering
 	\caption{Contour plot of the large sample coverage probability of the post-model-selection confidence interval, with nominal coverage 0.95, 
 	              for the odds ratio of MI in relation to recent OC use, for the \textbf{case control example}.		
 		   This confidence interval is constructed after a preliminary test of the null hypothesis that the coefficients of two specified second order interaction terms are both zero 
 		(i.e. $q = 2$).} 
 \end{figure}

\medskip

In Section 4 we use 
Theorem 1
to compare the finite sample coverage probability (found by simulation) of Schlesselman's (1982) post-model-selection confidence interval with its large sample coverage probability. 
We envisage two applications of our 
Theorems 1 and 2.
Firstly, they can be used to swiftly  provide a good indication of whether or not the post-model-selection confidence interval has minimum coverage well below its nominal coverage. 
Secondly, 
they can be used to swiftly narrow down the regions in the parameter space where one would search for the finite sample minimum coverage via simulation.

\bigskip

\noindent {\bf{2. An initial expression for the large		
sample coverage probability of a confidence interval obtained
after a hypothesis test concerning a vector parameter}}

\medskip

In this section we present the expression for the large sample coverage probability 
given directly below Figure 1 of  Hjort \& Claeskens (2003) for
the particular case of a preliminary hypothesis test concerning a vector parameter.
We consider a general regression model with response vector 
 $(y_1, \dots, y_n)$. 
 The random variables $y_1, \dots, y_n$ are independent
 and $y_i$ has density $f(y \, | \, \bx_i, \bphi)$, where the parameter vector
 $\bphi = (\btheta, \bgamma)$,
 with $\btheta$ a $p$-vector, $\bgamma$ a $q$-vector and
 $\bx_i$ a vector of explanatory variables of given dimension ($i=1, \dots, n$).
 Let $I(\btheta, \bgamma)$ denote the information matrix. In other words, let
 $I(\btheta, \bgamma) = E\left( - (\partial / \partial \bphi) (\partial / \partial \bphi)^T  \sum_{i=1}^n \log f(y_i \, | \, \bx_i ; \bphi) \right)$, 
 where
 $\partial / \partial \bphi$ 
 denotes the column vector of partial derivatives.
 We suppose, as do Hjort \& Claeskens (2003, p.883), that 
 $n^{-1} I(\btheta, \bgamma)$ converges to a finite nonsingular matrix as 
 $n \rightarrow \infty$, for each possible value of $(\btheta, \bgamma)$.
We also assume that the regularity conditions required for Lemmas 3.1--3.3 and Theorem 4.1
of  Hjort \& Claeskens (2003) to hold in the general regression framework
are satisfied (see Hjort \& Claeskens, 2003, p.884, Fahrmeir \& Kaufmann, 1985
and Fahrmeir \& Tutz, 1994, pp.43--44 
).
 
We also consider a restricted model that results from setting $\bgamma$
equal to the specified value  $\widetilde{\bgamma}$.
Suppose that the parameter of interest is $\varphi = \bs{a}^T\btheta$, where $\bs{a}$ is a specified non-zero $p$-vector.
Denote the maximum likelihood estimate (MLE) of $\bs{\phi}$ by 
$\widehat{\bs{\phi}} = (\widehat{\btheta}, \widehat{\bgamma})$. 
Also denote the MLE of $\btheta$ under the restricted model by $\widehat{\btheta}_r$.
The MLE's of $\varphi$ are $\widehat{\varphi} =  \bs{a}^T\widehat{\btheta}$ and $\widehat{\varphi}_r =  \bs{a}^T\widehat{\btheta}_r$ under the full and restricted models, respectively.
Partition the information matrix $I(\btheta, \bgamma)$ and its inverse as follows 
\begin{align*}
I(\btheta, \bgamma) = \begin{bmatrix} I_{\theta \theta}(\btheta, \bgamma)& I_{\theta \gamma}(\btheta, \bgamma)\\
					       I_{\gamma \theta}(\btheta, \bgamma)& I_{\gamma \gamma}(\btheta, \bgamma)
\end{bmatrix}
\, \text{ and } \, 
I^{-1}(\btheta, \bgamma) =
\begin{bmatrix} I^{\theta \theta}(\btheta, \bgamma) & I^{\theta \gamma}(\btheta, \bgamma)  \\
                          I^{\gamma \theta}(\btheta, \bgamma) & I^{\gamma \gamma}(\btheta, \bgamma)
\end{bmatrix}.
\end{align*}

Let $[a \pm b]$ denote the interval $[a-b, a+b]$ ($b \ge 0$).
Suppose that the confidence intervals for $\varphi$, with nominal coverage $1-\alpha$, are  
$J = \left[\widehat{\varphi} \pm z_{1-\alpha/2} (\bs{a}^{T}I^{\theta \theta}(\widehat{\btheta}, \widehat{\bgamma}) \bs{a})^{1/2} \right]$    
and 
$J_r = \left[ \widehat{\varphi}_{r} \pm z_{1-\alpha/2} (\bs{a}^{T} \big(\bs{I}_{\theta \theta}(\bs{\widehat{\theta}}_{r}, \widetilde{\bs{\gamma}})\big)^{-1}   \bs{a})^{1/2} \right]$ 
under the full and restricted models, respectively.
Here, $z_a$ denotes the inverse of the $N(0,1)$ cdf, evaluated at $a$.
Also suppose that we carry out a preliminary 
test of $H_0: \bs{\gamma} = \widetilde{\bs{\gamma}}$ against $H_A: \bs{\gamma} \neq \widetilde{\bs{\gamma}}$, rejecting
$H_0$ when the Wald test statistic 
$W = (\widehat{\bgamma} - \widetilde{\bs{\gamma}} )^{T} \big(I^{\gamma \gamma}(\widehat{\btheta}_{r}, \widetilde{\bs{\gamma}})\big)^{-1} (\widehat{\bgamma} - \widetilde{\bs{\gamma}} )$
exceeds $\chi^{2}_{1-\widetilde{\alpha}, q}$. 
Here, $\chi^{2}_{a, q}$ denotes the inverse of the $\chi^{2}_q$ cdf, evaluated at $a$. 
In other words, this test has large sample size $\widetilde{\alpha}$. 
The post-model-selection confidence interval $K$ for $\varphi$, with nominal coverage $1-\alpha$,  is defined as follows. If $H_0$ is accepted then $K = J_r$; otherwise $K = J$.

Define the $q$-vectors
$\bs{b} =  \big(I^{\gamma \gamma}(\btheta, \widetilde{\bs{\gamma}})\big)^{-1/2}I^{\gamma \theta }(\btheta, \widetilde{\bs{\gamma}})\bs{a}/ { (\bs{a}^{T}I^{\theta \theta}(\btheta, \widetilde{\bs{\gamma}}) \bs{a})^{1/2}}$ 
and \newline
$\blambda= \big(I^{\gamma \gamma}(\btheta, \widetilde{\bgamma})\big)^{-1/2} (\bgamma-\widetilde{\bgamma})$. 
Let the random variable $V_1$ and the random $q$-vector $\bs{H}$ have joint distribution
\begin{align}
\begin{bmatrix} 
V_1 \\
\bs{H}				
\end{bmatrix}
\sim
N
\left(
\begin{bmatrix}
0 \\
\blambda
\end{bmatrix}
,
\begin{bmatrix}
1 & \bs{b}^T \\
\bs{b} & \bs{I}_q 
\end{bmatrix}
\right).
\label{eq:V1Hjointdist}
\end{align}
Thus the distribution of $V_1$ conditional on $\bs{H} = \bs{h}$ is 
$N\left(\bs{b}^T(\bs{h}-\blambda), 1 - \|\bs{b}\|^2\right)$, 
where  $\| \,\cdot\, \|$ denotes the Euclidean norm.
Also let the random variable $V_2$ have distribution 
$ N\left(- \bs{b}^T \blambda/(1 - \|\bs{b}\|^2)^{1/2}, 1 \right)$, conditional on $\bs{H} = \bs{h}$.
For any statement $\mathcal{A}$, let $\mathcal{I}(\mathcal{A}) = 1$ if $\mathcal{A}$ is true; otherwise $\mathcal{I}(\mathcal{A}) = 0$.
For the scenario that we consider, the expression for the large sample coverage probability given directly below Figure 1 of Hjort \& Claeskens (2003) is, 
as shown in the Supporting Information,
the following. 
As $n \rightarrow \infty$, the coverage probability $P(\varphi \in K)$ approaches
\begin{align}
\begin{aligned}
&\int P\left(\left| V_2 \right| \leq z_{1-\alpha/2}\Big| \bs{H} = \bs{h} \right)  \mathcal{I}\left(\|\bs{h}\|^2 \leq  \chi^{2}_{1-\widetilde{\alpha}, q}\right)f_{\bs{H}}(\bs{h})d\bs{h}\\
&+ \int P\left(\left| V_1 \right| \leq z_{1-\alpha/2}\Big| \bs{H} = \bs{h} \right)  \mathcal{I}\left(\|\bs{h}\|^2 >  \chi^{2}_{1-\widetilde{\alpha}, q}\right)f_{\bs{H}}(\bs{h})d\bs{h}, 
\label{eq:meetingpointnew}
\end{aligned}
\end{align}
where $f_{\bs{H}}$ denotes the pdf of $\bs{H}$. 
This result may also be obtained using a straightforward extension to the local misspecification framework of equation $(5)$ in Section 2 of Cox \& Wermuth (1990, p.748).

\bigskip

\noindent {\bf{3. The main result}}

\medskip

Our main result is 
Theorem 1
which states that the expression \eqref{eq:meetingpointnew} is equal to a formula consisting of a trivial term added to either
a double integral for $q = 2$ or a triple integral for all $q > 2$. 
These integrals are readily evaluated by numerical integration.
The proof of Theorem 1 is given in the appendix and uses the methodology in the appendix of
Kabaila \& Farchione (2012). This methodology consists of the following components.

\begin{enumerate}
	
	\item[1.] 
	
	For $q \ge 2$, we express a random $q$-vector with an $N(\bzero, \bs{I}_q)$ distribution
	as $R \, \bs{U}$, where $\bs{U}$ and $R$ are independent, with $\bs{U}$ uniformly distributed on the surface of the unit sphere in $\mathbb{R}^q$ and 
	$R \sim \chi_q$ (so that $R^2 \sim \chi^2_q$).
	
	\item[2.] 
	
	We note that for any unit vectors $\bs{u}$ and $\widetilde{\bs{u}}$ in $\mathbb{R}^q$ the following are true.
	
	\begin{enumerate}
		
		\item[(a)]
		
		The inner product $\bs{u}^T \, \bs{U}$ has a distribution that does not
		depend on the orientation of the vector $\bs{u}$ and consequently has 
		the same distribution as $U_1$.
		
		\item[(b)] 
		
		For given $q \ge 2$, the distribution of the random vector
		$\big(\bs{u}^T \, \bs{U}, \widetilde{\bs{u}}^T \, \bs{U} \big)$
		depends only on the inner product 
		$\psi = \bs{u}^T \, \widetilde{\bs{u}}$. 
		Indeed, $\big(\bs{u}^T \, \bs{U}, \widetilde{\bs{u}}^T \, \bs{U} \big)$
		has the same distribution as 
		$\big(U_1, \psi \, U_1 + (1 - \psi^2)^{1/2} \, U_2\big)$.
		
		\item[(c)] 
			
		The spherical coordinate transformation, stated e.g. on p.306 of
		Fang \& Wang (1994), can be used to express $U_1$ and $U_2$ in terms
		of the independent random variables $T_1$ and $T_2$ with pdf's given
		by \eqref{eq:Pdf_T1} and \eqref{eq:Pdf_T2}, respectively.
		 
	\end{enumerate}
	
\end{enumerate}

\noindent 
Components 1 and 2(c) of this methodology lead to the presence of the
pdf's $f_{T_1}$, $f_{T_2}$ and the pdf $f_R$ of $R$ in the formula for the large sample
coverage probability given in the following theorem. The component 2(c) leads
to the presence of the sin and cos terms in this formula. Finally, component 2(b) leads 
to this formula being a trivial term added to a triple integral for all $q > 2$. 
The following is our main result.

\begin{theorem}
\label{thm1mod}

For all $\bgamma = \widetilde{\bgamma} + n^{-1/2} \bs{\delta}$ $(\bs{\delta} \in [-d, d]^q, \,  0 < d < \infty)$,  
the coverage probability $P(\varphi \in K)$ approaches
\begin{align}
P\left( |V_2| \leq z_{1-\alpha/2} \right) P(\|\bs{H}\|^2 \leq  \chi^{2}_{1-\widetilde{\alpha}, q})
+ B_q
\label{eq:final}
\end{align}
as $n \rightarrow \infty$. 
Here 
$V_2 \sim N\left(- \psi \|\bs{b}\| \|\blambda\| /(1 - \|\bs{b}\|^2)^{1/2}, 1 \right)$, 
where 
$\psi =  (\bs{b}/ \|\bs{b} \| )^{T} ( \blambda/ \|\blambda\|)$
for $\bs{b}\ne \bzero$ and $\blambda \ne \bzero $ (otherwise $\psi = 1$),
and $B_q$ is defined as follows.
Let 
$i( \upsilon ; \|\bs{b} \| ) = 
P\left(
-z_{1-\alpha/2}  \leq V_3 + \upsilon \leq z_{1 - \alpha/2}   
\right)$,
where
$V_3 \sim N(0, 1 - \|\bs{b}\|^2)$ and $\upsilon \in \mathbb{R}$.
Also let
$k(t_{1}, \psi)  = \psi \cos(2 \pi t_{1}) + (1 - \psi^{2})^{1/2} \sin(2 \pi t_{1})$.
Then
\begin{align}
 B_2 
 &= 
 1 - \alpha - \int\displaylimits^{1}_{0}\int\displaylimits_{[l_2, u_2]\cap [0, \infty)} i\big(r \, \|\bs{b} \| \, k(t_{1}, \psi) \, ; \, \|\bs{b} \| \big) \, f_R(r) \, dr \, dt_1,   
 \label{eq:rq2}
 \end{align}
where 
$[l_2,  u_2] =  \left[- \|\blambda\| \cos(2 \pi t_{1}) \pm \left(\|\blambda\|^{2}\cos^{2}(2 \pi t_{1}) + \chi^{2}_{1-\widetilde{\alpha}, 2}  - \|\blambda\|^{2} \right)^{1/2}  \right]$
and $f_R$ denotes the $\chi_q$ pdf.
Let 
$k(t_{1}, t_{2}, \psi; 3) = \psi\cos(\pi t_{1}) + (1 - \psi^{2})^{1/2} \sin( \pi t_{1})\cos(2 \pi t_{2})$ and 
$k(t_{1}, t_{2}, \psi; q) =\psi\cos(\pi t_{1}) + (1 - \psi^{2})^{1/2} \sin( \pi t_{1})\cos(\pi t_{2})$ for $q >3$.
Then, for $q>2$
\begin{align}
B_q
& = 
1 - \alpha -  \int\displaylimits^{1}_{0}\int\displaylimits^{1}_{0} \int\displaylimits_{[l_q, u_q]\cap [0, \infty)} i \big(r \, \|\bs{b} \| \,  k(t_{1}, t_{2}, \psi ; q) ;\|\bs{b} \| \big) \, f_R(r) \, dr  \, f_{T_{1}}(t_{1}) \, dt_{1} \, f_{T_{2}}(t_{2}) \,  dt_{2}, 
 \end{align}
where 
$[l_q,  u_q] =  \left[- \|\blambda\| \cos( \pi t_{1}) \pm \left( \|\blambda\|^{2}\cos^{2}( \pi t_{1}) + \chi^{2}_{1-\widetilde{\alpha}, q}  - \|\blambda\|^{2} \right)^{1/2} \right]$,
\begin{align}
\label{eq:Pdf_T1}
f_{T_{1}}(t_{1}) &=
\displaystyle \frac{\pi \sin^{q-2}(\pi t_{1})}{B\left(1/2, (q-1)/2\right)}  \ ,0 \le t_1 \le 1, 
 \, q \ge 2  
\quad \text{ and } \quad 
\\
\label{eq:Pdf_T2}
f_{T_{2}}(t_{2}) &=
\displaystyle \frac{\pi  \sin^{q-3}(\pi t_{2}) }{B\left(1/2, (q-2)/2\right)}  \ ,0 \le t_2 \le 1, 
 \, q > 2,
\end{align}
with $B(a, b)$ denoting the beta function.
When $\|\bs{b}\| > 0$ and 
 $\blambda = \bs{0}$, \eqref{eq:final} simplifies to 
\begin{align}
(1-\alpha)(2-\widetilde{\alpha}) - \int_0^1 \int_0^{(\chi_{1 - \widetilde{\alpha}, q}^2)^{1/2}} i\left(r \, \|\bs{b}\| \, g(t_1;q); \, \|\bs{b}\| \, \right) \, f_R(r) \, dr  \, f_{T_1}(t_1)\, dt_1,
\label{LargeSampleCovNormbGT0Lambda0}
\end{align}
where $g(t_1 ; 2) = \cos(2\pi t_1)$ and $g(t_1 ; q) = \cos(\pi t_1)$ for $q >2$.
For $\bs{b} = \bs{0}$,  \eqref{eq:final} simplifies to $1 - \alpha$.
\end{theorem}

This theorem has two appealing properties. 
Firstly, the large sample coverage probability
(given by \eqref{eq:final}) 
requires the evaluation of at most a triple integral,  irrespective of the dimension $q$ of $\bgamma$. 
Secondly, this coverage probability is, for given $\btheta$, a function of two unknown scalar values, namely $\psi$ and $\|\blambda\|$
 irrespective of the dimension $q$,
 and three known quantities,  namely
 the nominal coverage $1 - \alpha$, the nominal level of significance $\widetilde{\alpha}$ and $\|\bs{b}\|$. 
As a result, the large sample coverage probability, minimized over  
 $(\|\blambda\|,\psi)$, can be easily computed for given $\btheta$ and given values of these known quantities.

The following theorem leads to a halving of the time required to compute the large sample coverage probability, minimized over $(\|\blambda\|,\psi)$.
\begin{theorem}
\label{thm2}

Suppose that $\widetilde{\bgamma} = \bs{0}$
and that  $1-\alpha$, $\widetilde{\alpha}$ and $\|\bs{b}\|$
are given. For given  $\btheta$, the large sample coverage probability
 \eqref{eq:final}  evaluated at $\bgamma = \bgamma^{\prime}$  is equal to 
 \eqref{eq:final}
 evaluated at $\bgamma = -\bgamma^{\prime}$, for all
 $\bgamma^{\prime}$. In other words, this coverage probability 
 is an even function of $\psi \in [-1, 1]$.
\end{theorem}

Let $\text{LSCP} \big(\|\bs{b} (\btheta)\|,\|\blambda\|, \psi \big)$ denote the large sample coverage probability \eqref{eq:final},
where the dependence of $\bs{b}$ on $\btheta$ is made explicit in the notation $\bs{b} (\btheta)$. 
A summary description of this large sample coverage probability function is 
\begin{equation}
\label{IdealMinCov}
\min_{\|\blambda\|, \psi} \text{LSCP} \big(\|\bs{b} (\btheta)\|,\|\blambda\|, \psi \big),
\end{equation}
where $\btheta$ denotes the true parameter value. 
Here we make a sharp distinction between the parameter vectors $\btheta$ and $\bgamma$.
We estimate \eqref{IdealMinCov}, for a particular data set, by
\begin{equation}
\label{EstMinCov}
\min_{\|\blambda\|, \psi} \text{LSCP} \big(\|\bs{b}(\widehat{\btheta}_{obs})\|,\|\blambda\|, \psi \big),
\end{equation}
where $\widehat{\btheta}_{obs}$ denotes the maximum likelihood estimate
of $\btheta$ based on this data set.
In other words, we use a ``plug-in principle'' (see e.g. Efron, 1998, Section 5) approach for this parameter.
However, we do not replace $\bgamma$ by an estimate because of the local misspecification framework that we must use for this parameter. 
This leads to the parameters $\|\blambda\|$ and  $\psi$ also not being replaced by estimates. 
A conceptually similar approach has been used by Kabaila, Mainzer \& Farchione (2017).
Therefore, when preparing Figure 1, the contour plot of the large sample coverage probability of the post-model selection confidence interval for the \textbf{case control example}, we have replaced $\btheta$ by its maximum likelihood estimate $\widehat{\btheta}_{obs}$ based on this data set.
Note that this figure provides an illustration of 
Theorem 2.

It is expected that, for large sample sizes, the difference between
\eqref{IdealMinCov} and \eqref{EstMinCov}
will be small.
We propose the following parametric bootstrap method to assess this difference. 
We set the true parameter value $(\btheta, \bgamma)$ equal to its
maximum likelihood estimate
$(\widehat{\btheta}_{obs}, \widehat{\bgamma}_{obs})$. 
We then generate $B$ independent observations of the response vector.  
For the $b$'th of these observations we compute the maximum likelihood estimate 
$\widehat{\btheta}_b^*$
and then replace $\widehat{\btheta}_{obs}$ by $\widehat{\btheta}_b^*$
in \eqref{EstMinCov} to obtain a parametric bootstrap resample of \eqref{EstMinCov}. 
These bootstrap resamples of  \eqref{EstMinCov} are then used to construct a confidence interval for \eqref{IdealMinCov}.
We applied this method, with $B = 1000$, to the post-model-selection confidence interval, with nominal coverage 0.95,  in the \textbf{case control example}. 
We obtained the $95\%$ percentile interval $[0.2120, 0.4221]$ for \eqref{IdealMinCov}. 
We also obtained the $95\%$ bootstrap confidence interval $[0.2341, 4441]$ for \eqref{IdealMinCov}, using (13.9) of Efron \& Tibshirani (1993).
 Both of these intervals suggest that for the post-model-selection confidence interval found by Schlesselman (1982) the value of \eqref{IdealMinCov}
is far below the nominal coverage, indicating that this post-model-selection confidence interval should not be used.

\medskip

\noindent {\bf{4. A comparison of the large sample and finite sample coverage probabilities}}

\medskip
		
To assess the accuracy of the large sample approximation \eqref{eq:final}
to the finite sample coverage probability of the post-model-selection confidence interval $K$, we compare these coverage probabilities 
in the \textbf{case control example} as functions of $\bs{\gamma}$, for $\btheta$ replaced by its maximum likelihood estimate $\widehat{\btheta}_{obs}$. 
Specifically, we compare these coverage probabilities as functions of 
$\gamma_{both} = \gamma_1 = \gamma_2$, with $\widetilde{\bs{\gamma}} = \bs{0}$, $1-\alpha = 0.95$ and $\widetilde{\alpha} = 0.05$.  
The finite sample coverage probability is estimated by simulation,
 with $40000$ simulations for each value of $\gamma_{both}$ considered. 
The large sample coverage probability is found using \eqref{eq:final} and \eqref{eq:rq2} of 
Theorem 1.
A detailed description of the data for the \textbf{case control example} is included in the Supporting Information. 
For this data, the length of the response vector is $n = 1976$.

Figure 2 shows graphs of the large sample and finite sample coverage probabilities of this post-model-selection confidence interval.
The positive  values of $\gamma_{both}$ for which these coverage probability functions are minimized are close. 
Also, the negative values of $\gamma_{both}$ for which these coverage probability functions are minimized are close. 
Furthermore, the minima over $\gamma_{both}$ of these coverage probability functions are also close.
\begin{figure}[H]
	\includegraphics[width = 0.75\textwidth]{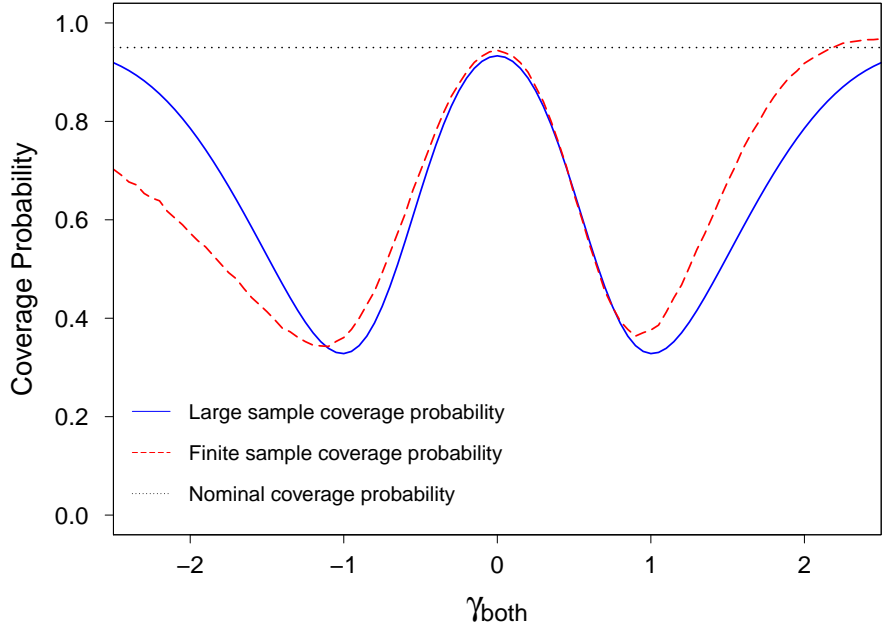}
\centering
\caption{Comparison of the large sample and finite sample coverage probabilities of the  post-model-selection confidence interval  for the \textbf{case control example}.}
\end{figure}
The 40000 simulations used for each value of $\gamma_{both}$ led to an estimator of the finite sample coverage probability with standard deviation guaranteed to be less than or equal to 0.0025. 
In Figure 2 we used 101 equally-spaced values of $\gamma_{both}$. The time taken to compute the finite sample coverage probabilities plotted in Figure 2 was about 5.5 hours on a PC with an Intel Core i7-4790, 3.60GHz CPU and 16GB of RAM. 
Using this computer, the time taken to compute the large sample coverage probabilities  plotted in this figure was about 3.5 seconds i.e. smaller by a factor of over 5000. 

To demonstrate that the large sample and finite sample coverage probabilities become closer as the sample size is increased, 
we considered $128$ independent replications (with the same values of the explanatory variables) of the experiment that gave rise to the \textbf{case control example} data. 
Graphs of the resulting large sample and finite sample coverage probability functions are shown in Figure 3.  
In Figure 3 we used 101 equally-spaced values of $\gamma_{both}$.
\begin{figure}[H]
	\includegraphics[width = .75\textwidth]{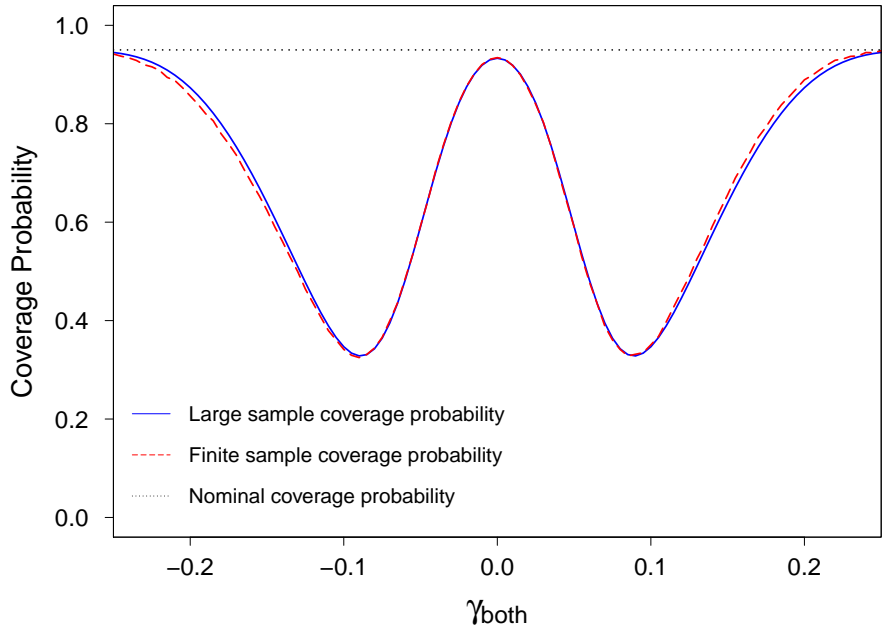}
\centering
\caption{Comparison of the large sample and finite sample coverage probabilities of the  post-model-selection confidence interval for $128$ independent replications of the experiment that gave rise to the \textbf{case control example} data.
}
\end{figure}

\noindent {\bf{5. Conclusion}}

\medskip

In this paper we consider a general regression model specified by the $p$-vector $\btheta$ and the $q$-vector $\bgamma$. 
We suppose that a preliminary hypothesis test, with large sample size $\widetilde{\alpha}$, is used to select one of
two nested models: 
the full model and a restricted model in which $\bs{\gamma}$ takes the specified value $\widetilde{\bs{\gamma}}$. 
We have derived a new computationally convenient formula \eqref{eq:final} for the large sample coverage probability of the post-model-selection confidence interval, with nominal
coverage $1 - \alpha$, for a scalar parameter of interest. 
This formula consists of a trivial term added to either a double integral for $q = 2$ or a triple integral for all $q > 2$. 
 These multiple integrals are readily evaluated by numerical integration.

Theorems 1 and 2
imply that, for given $\widetilde{\alpha}$, $1 - \alpha$ and $\btheta$, 
this large sample coverage probability is a function of only two scalar parameters, $\psi \in [0, 1]$ and $\|\blambda\|$, for all $q \ge 2$. 
We can therefore quickly compute the large sample coverage probability minimized with respect to these two parameters.
Theorem 1 also provides us with the insight that $\btheta$ influences this minimized coverage probability only through the scalar parameter $\|\bs{b}\|$.

We have put forward the following procedure for the rapid assessment of the coverage properties of the post-model-selection confidence interval. 
Firstly, we replace $\btheta$ by its maximum likelihood estimate $\widehat{\btheta}_{obs}$. This is a ``plug-in principle'' 
approach.  
We then minimize the large sample coverage probability \eqref{eq:final} with respect to the two scalar parameters $\psi$ and $\|\blambda\|$. 
We have made a sharp distinction between the parameter vector $\btheta$ and these two scalar parameters, which we do not replace by estimates, because of the local misspecification framework that we must use for the parameter $q$-vector $\bgamma$. 
For the \textbf{case control example} data, the large sample coverage minimized in this
way is close to the minimum finite sample coverage evaluated using simulations, which are relatively very time-consuming. 
Finally, we assess the impact of replacing $\btheta$ by $\widehat{\btheta}_{obs}$ using standard parametric bootstrap methodology. 

\bigskip

\noindent {\bf{Acknowledgements}}

This work was supported by an Australian Government Research Training Program Scholarship. 

\medskip

\noindent {\bf{Supporting Information}}

\noindent Additional information for this article is available online.

\noindent Description: Additional derivations, proofs and descriptions.

\medskip

\noindent {\bf{References}} 

\medskip

\rf Bendas, A., Rothe, U., Kiess, W., Kapellen, T. M., Stange, T., Manuwald, U., Salzsieder, E., Holl, R. W., Schoffer, O., Stahl-Pehe, A., Giani, G., Ehehalt, S., Neu, A. \& Rosenbauer, J. (2015). Trends in incidence rates during 1999-2008 and prevalence in 2008 of childhood type 1 diabetes mellitus in GERMANY - model-based national estimates.  \textit{PLoS ONE}  {\bf{10}}, 1--12.

\rf Cox, D. R. \& Barndorff-Nielsen, O. E. (1989). \textit{Asymptotic Techniques for use in Statistics}. Chapman \& Hall, London.

\rf Cox, D. R. \& Wermuth, N. (1990). An approximation to maximum likelihood estimates in reduced models. \textit{Biometrika} {\bf{77}},  
747--761.

\rf Efron, B. (1998). R.A. Fisher in the 21st century. \textit{Stat. Sci.}
\textbf{13}, 95--112.

\rf Efron, B. \& Tibshirani, R. J. (1993). \textit{An Introduction to the Bootstrap}. Chapman \& Hall, London.

\rf Fahrmeir, L. \& Kaufmann, H. (1985). Consistency and asymptotic normality of the 
maximum likelihood estimator in generalized linear models. \textit{Ann. Statist.}
\textbf{13}, 342--368.

\rf Fahrmeir, L. \& Tutz, Z. (1994). \textit{Multivariate Statistical Modelling Based on 
Generalized Linear Models}. Springer-Verlag, New York.

\rf Fang, K. -T. \& Wang, Y. (1994). \textit{Number-Theoretic Methods in Statistics}. Chapman \& Hall, London.

\rf Hjort, N. L. \& Claeskens, G. (2003). Frequentist model average estimators. \textit{J. Amer. Statist. Assoc.} {\bf{98}}, 
879--899.

\rf Huber, M., Andersohn, F., Bronder, E., Klimpel, A., Thomae, M., Konzen, C., Meyer, O., Salama, A., Schrezenmeier, H., Hildebrandt, M., Sp\"ath-Schwalbe, E., Gr\"uneisen, A., Kreutz, R. \& Garbe, E. (2014). Drug-induced agranulocytosis in the Berlin case-control surveillance study. \textit{Eur. J. Clin. Pharmacol.} {\bf{70}}, 339--345.

\rf Kabaila, P. (2009). The coverage properties of confidence regions after model selection. \textit{Int. Statist. Review} {\bf{77}}, 405--414.

\rf Kabaila, P. \& Farchione, D. (2012). The minimum coverage probability of confidence intervals in regression after
a preliminary F test. \textit{J. Stat. Plan. Inf.} {\bf{142}}, 956--964.

\rf Kabaila, P., Mainzer, R. \& Farchione, D. (2017). Conditional assessment of the impact of a Hausman pretest on confidence intervals.  \textit{Stat. Neerl.}
doi:10.1111/stan.12109.

\rf Kabat, G. C., Jones, J. G., Olson, N., Negassa, A., Duggan, C., Ginsberg, M., Kandel, R. A., Glass, A. G. \& Rohan, T. E. (2010). A multi-center prospective cohort study of benign breast disease and risk of subsequent breast cancer. \textit{Cancer Causes Control} {\bf{21}}, 821--828.

\rf  Kanbayashi, Y. Matsumoto, Y., Kuroda, J., Kobayashi, T., Horiike, S., Hosokawa, T. \& Taniwaki, M. (2017). Predicting risk factors for varicella zoster virus infection and postherpetic neuralgia after hematopoietic cell transplantation using ordered logistic regression analysis, \textit{Ann. Hematol.} {\bf{96}}, 311--315.

\rf Li, D., Tang, H., Hassan, M. M., Holly, E. A., Bracci, P. M. \& Silverman, D. T. (2011). Diabetes and risk of pancreatic cancer: a pooled analysis of three large case-control studies. \textit{Cancer Causes Control} {\bf{22}}, 189--197.

\rf McCullagh, P. \& Nelder, J. A. (1989). \textit{Generalized Linear Models}, 2nd edn. Chapman \& Hall, London.

\rf Nelder, J. A. \& Wedderburn, R. W. M. (1972). Generalized linear models.  \textit{J. R. Stat. Soc.: Series A} {\bf{135}}, 370--384.

\rf O'Donnell, M. J., Xavier, D., Liu, L., Zhang, H., Chin, S. L., Rao-Melacini, P., Rangarajan, S., Islam, S., Pais, P., McQueen, M. J., Mondo, C., Damasceno, A., Lopez-Jaramillo, P., Hankey, G. J., Dans, A. L., Yusoff, K., Truelsen, T., Diener, H., Sacco, R. L., Ryglewicz, D., Czlonkowska, A., Weimar, C., Wang, X. and Yusuf, S. (2010). Risk factors for ischaemic and intracerebral haemorrhagic stroke in $22$ countries (the INTERSTROKE study): a case-control study. \textit{Lancet} {\bf{376}}, 112--123.

\rf Schlesselman, J. J. (1982). \textit{Case-Control Studies}. Oxford University Press, New York.

\rf Stampf, S., Graf, E., Schmoor, C. \& Schumacher, M. (2010). Estimators and confidence intervals for the marginal odds ratio using logistic regresson and propensity score stratification. \textit{Stat. Med.} {\bf{29}}, 760--769.

\bigskip

\noindent {\bf{Appendix}}

\medskip

\noindent {\it{Proof of Theorem 1}}
 
 \medskip

Let the first and second multiple integrals in the expression \eqref{eq:meetingpointnew} be denoted by $A$ and $B_q$, respectively. In other words, let 
\begin{align*}
A = \displaystyle \int P\left(\left| V_2 \right| \leq z_{1-\alpha/2}\Big| \bs{H} = \bs{h} \right)  \mathcal{I}\left(\|\bs{h}\|^2 \leq  \chi^{2}_{1-\widetilde{\alpha}, q}\right)f_{\bs{H}}(\bs{h})d\bs{h} 
\end{align*}
and 
\begin{align*}
B_q = \displaystyle \int P\left(\left| V_1 \right| \leq z_{1-\alpha/2}\Big| \bs{H} = \bs{h} \right)  \mathcal{I}\left(\|\bs{h}\|^2 >  \chi^{2}_{1-\widetilde{\alpha}, q}\right)f_{\bs{H}}(\bs{h})d\bs{h}.
\end{align*}
Note that as $V_2$ and $\bs{H}$ are independent, 
\begin{align}
A
& = P\left(\left| V_2 \right| \leq z_{1-\alpha/2}\right)  P\left(\|\bs{H}\|^2 \leq  \chi^{2}_{1-\widetilde{\alpha}, q} \right).
\label{eq:simplea}
\end{align}
Now 
\begin{align}
B_q = P(| V_1 | \leq z_{1-\alpha/2},
\|\bs{H}\|^2 >  \chi^{2}_{1-\widetilde{\alpha}, q}).
\label{eq:bqjoint}
\end{align}
By the law of total probability,
\begin{align*}
P \big(| V_1 | \leq z_{1-\alpha/2} \big)
&= P \big(| V_1 | \leq z_{1-\alpha/2},
\|\bs{H}\|^2 >  \chi^{2}_{1-\widetilde{\alpha}, q} \big)
+ P \big(| V_1 | \leq z_{1-\alpha/2},
\|\bs{H}\|^2 \le  \chi^{2}_{1-\widetilde{\alpha}, q} \big).
\end{align*}

It follows from \eqref{eq:V1Hjointdist}
 that $V_1 \sim N(0,1)$. Therefore, 
$P \big(| V_1 | \leq z_{1-\alpha/2} \big) = 1 - \alpha$. Hence
\begin{align}
B_q
 = 
1 - \alpha - \int P\left(\left| V_1 \right| \leq z_{1-\alpha/2}\Big| \bs{H} = \bs{h} \right)  \mathcal{I}\left(\|\bs{h}\|^2 \leq \chi^{2}_{1-\widetilde{\alpha}, q}\right)f_{\bs{H}}(\bs{h}) \, d\bs{h}. 
\end{align}
Note that
\begin{align*}
 P\left(\left| V_1 \right| \leq z_{1-\alpha/2} \, \Big|\, \bs{H} = \bs{h} \right) 
  & = P\left(- z_{1-\alpha/2}  - \bs{b}^T(\bs{h} - \blambda) \leq V_3 \leq z_{1-\alpha/2}  - \bs{b}^T(\bs{h} - \blambda) \right) \\
  & = i \big( \bs{b}^T(\bs{h} - \blambda); \|\bs{b}\| \big).
\end{align*}
Therefore
\begin{align}
B_q
& = 1 - \alpha - E\Big( i(\bs{b}^T(\bs{H} - \blambda); \|\bs{b}\|) \, \mathcal{I}\left(\|\bs{H}\|^2 \leq  \chi^{2}_{1-\widetilde{\alpha}, q}\right)\Big).
\label{eq:bq}
\end{align}

For the moment, consider the case that $\bs{b} = \bs{0}$. In this case, $V_2 \sim N(0,1)$ and $V_3 \sim N(0,1)$, so that 
$A =(1 - \alpha) P\left(\|\bs{H}\|^2 \leq  \chi^{2}_{1-\widetilde{\alpha}, q} \right)$ 
and 
$B_q = 1 - \alpha - (1-\alpha) P\left(\|\bs{H}\|^2 \leq  \chi^{2}_{1-\widetilde{\alpha}, q} \right)$. 
Therefore, $A + B_q = 1 - \alpha$.

We now consider the case that $\|\bs{b}\| > 0$, and apply the methodology briefly 
outlined immediately before the statement of Theorem 1, 
 to \eqref{eq:bq} to obtain the large sample coverage probability formula \eqref{eq:final} of Theorem 1. 
Since $\bs{H} \sim N(\blambda, \bs{I}_q)$, we may write $\bs{H} = \blambda + R \, \bs{U}$,
where $\bs{U}$ and $R$ are independent, with $\bs{U}$ uniformly distributed on the surface of the unit sphere in $\mathbb{R}^q$ and
$R \sim \chi_q$ (so that $R^2 \sim \chi^2_q$).
Define the unit length vector $\bs{u}_{\bs{b}} = \bs{b}/\|\bs{b} \|$ and let
$L_{\bs{b}} =\bs{u}_{\bs{b}}^{T} \, \bs{U}$. 

There are two subcases: $\blambda = \bs{0}$ and $\| \blambda \| > 0$. We first consider the subcase $\blambda = \bs{0}$.
It follows from \eqref{eq:bq} that
\begin{align*}
B_q
 = 1 - \alpha - E\Big( i\left(\bs{b}^T \bs{H}; \|\bs{b}\| \right) \, \mathcal{I}\left(\|\bs{H}\|^2 \leq  \chi^{2}_{1-\widetilde{\alpha}, q}\right)\Big).
\end{align*}
Since  $\bs{b}^T \bs{H} = R \, \|\bs{b}\|\, L_{\bs{b}}$ and $\|\bs{H}\|^2 = R^2$,
\begin{align*}
B_q
= 1 - \alpha - E\Big( i\left(R \, \|\bs{b}\|\, L_{\bs{b}}; \|\bs{b}\| \right) \, \mathcal{I}\left(R^2 \leq  \chi^{2}_{1-\widetilde{\alpha}, q}\right)\Big).
\end{align*}
Define the random variables $T_1$ and $T_2$ to be such that 
$T_1, T_2$ and $R$ are independent and $T_1$ and $T_2$ have 
pdf's $f_{T_1}$ and  $f_{T_2}$, respectively. 
Let $\bs{e}$ denote the unit length $q$-vector $(1, 0, \cdots , 0)$. 
Observe that  $L_{\bs{b}} =\bs{u}_{\bs{b}}^{T} \, \bs{U}$ has the same distribution as $\bs{e}^T \bs{U} = U_1$,
the first component of $\bs{U}$.  
Recall the definitions
$g(t_1; 2) = \cos(2 \pi t_1)$ and $g(t_1; q) = \cos(\pi t_1)$ for $q > 2$. 
As shown by Fang \& Wang (1994, p.49 and pp.305-308) using a spherical coordinate transformation, $U_1$ has the same distribution as $g(T_1; q)$. 
Therefore 
\begin{align*}
B_q  &= 1 - \alpha - \int_0^1 \int_0^{\infty} i \left(r \, \|\bs{b}\|\, g(t_1; q); \|\bs{b} \| \right) \, \mathcal{I}\left(r^2 \leq  \chi^{2}_{1-\widetilde{\alpha}, q}\right)f_{R}(r)  \, dr
\, f_{T_1}(t_1) \, dt_1
\\ 
 &= 1 - \alpha - \int_0^1 \int_0^{\left(\chi_{1 - \widetilde{\alpha}, q}^2\right)^{1/2}} i \left(r \, \|\bs{b}\|\, g(t_1; q); \|\bs{b} \| \right)f_{R}(r) \, dr \, f_{T_1}(t_1) \, dt_1.
\end{align*}
Since $A = (1-\alpha)(1-\widetilde{\alpha})$, \eqref{LargeSampleCovNormbGT0Lambda0} is true.

Now consider the subcase  $\| \blambda \| > 0$. 
Define the unit length $q$-vector $\bs{u}_{\blambda} = \blambda / \|\blambda\|$ and let $L_{\blambda} =  \bs{u}_{\blambda}^T \, \bs{U}$. It follows from \eqref{eq:bq} that
\begin{align}
B_q
= 1 - \alpha - E\Big( i( R \|\bs{b}\|\, L_{\bs{b}}; \|\bs{b}\|) \, \mathcal{I}\left( R^2 + 2RL_{\blambda}\|\blambda\|  + \|\blambda\|^{2} \leq  \chi^{2}_{1-\widetilde{\alpha}, q}\right)\Big), 
\label{eq:bqfinalcase}
\end{align}
since $\bs{b}^T \left(\bs{H}-\blambda \right) = R \, \|\bs{b}\|\, L_{\bs{b}}$
and 
$\|\bs{H}\|^2 
= (\blambda + R\bs{U})^{T}(\blambda + R\bs{U}) 
= R^2 + 2RL_{\blambda}\|\blambda\|  + \|\blambda\|^{2}$.
Define the  unit length $q$-vectors  $\bs{e_{\lambda}} = \left(1, 0, \cdots , 0\right)$
and 
$\bs{e_b} = \left(\psi, (1- \psi^2)^{1/2}, 0, \cdots , 0 \right)$, 
where $\psi =\bs{u}_{\bs{b}}^T \bs{u}_{\blambda}$. 
Observe that $\left( L_{\blambda}, \, L_{\bs{b}}\right)$ has the same distribution as 
$\left(\bs{e_\lambda}^T\bs{U} , \, \bs{e_b}^T\bs{U} \right) = \left(U_1, \, \psi \, U_1 + (1- \psi^2)^{1/2} \,  U_2 \right)$. 

As shown by Fang \& Wang (1994, p.49 and pp.305-308) using a spherical coordinate transformation, 
$(U_1, U_2)$ has the same distribution as $\big(\cos(2 \pi T_1), \, \sin(2 \pi T_1)\big)$
for $q=2$,
 $\big(\cos(\pi T_1), \, \sin(\pi T_1) \cos(2 \pi T_2)\big)$ 
 for $q=3$ and 
$\big(\cos(\pi T_1), \, \sin(\pi T_1) \cos(\pi T_2)\big)$ 
for $q >3$. 
Therefore $\left(L_{\blambda}, \,  L_{\bs{b}} \right)$ has the same distribution 
as 
$\left(\cos(2 \pi T_1), \, \psi \, \cos(2 \pi T_1) + (1- \psi^2)^{1/2} \, \sin(2 \pi T_1) \right)$ 
for $q = 2$, 
$\left(\cos( \pi T_1), \, \psi \, \cos( \pi T_1)+ (1- \psi^2)^{1/2} \,  \sin(\pi T_1) \cos(2 \pi T_2) \right)$ 
for $q = 3$ and \newline 
$\left(\cos( \pi T_1), \, \psi \, \cos( \pi T_1) + (1- \psi^2)^{1/2} \, \sin(\pi T_1) \cos(\pi T_2) \right)$ 
for $q >3$.
In other words,
$\left(L_{\blambda}, \,  L_{\bs{b}} \right)$ has the same distribution as
$\left(\cos(2\pi T_1), \, k(T_1, \psi) \right)$ for $q = 2$, and 
$\left(\cos(\pi T_1), \, k(T_1, T_2, \psi; q) \right)$ for $q > 2$. 
Hence $\|\bs{H}\|^2 = d(T_{1}, R; q, ||\blambda||)$, where
$d(t_{1}, r; 2, ||\blambda||) = r^{2} + 2||\blambda|| r \cos(2 \pi t_{1})  + ||\blambda||^{2}$ and $d(t_{1}, r; q, ||\blambda||) = r^{2} + 2||\blambda|| r \cos(\pi t_{1})+ ||\blambda||^{2}$ for $q > 2$.
It follows from \eqref{eq:bqfinalcase} that
\begin{align*}
B_q
= 
\begin{cases}
1 - \alpha - E\Big( i(R\|\bs{b}\|k(T_1, \psi); \|\bs{b}\|) \, \mathcal{I}\left( d(T_1, R; \, 2, \|\blambda\|) \leq  \chi^{2}_{1-\widetilde{\alpha}, 2}\right)\Big) 
&  \quad \,  \text{ for $q = 2$, }  \\ 
1 - \alpha - E\Big( i(R\|\bs{b}\|k(T_1, T_2, \psi); \|\bs{b}\|) \, \mathcal{I}\left( d(T_1, R; \, q, \|\blambda\|) \leq  \chi^{2}_{1-\widetilde{\alpha}, q}\right)\Big) 
& \quad \,  \text { for $q > 2$. }
\end{cases}
\end{align*}
%
Let 
$S_q(t_1) = \left\{r: d(t_{1}, r; q, \|\blambda\|) \leq  \chi^{2}_{1-\widetilde{\alpha}, q} \right \}$.
Since $r \in [0, \infty)$, 
\begin{align*}
B_q
= 
\begin{dcases}
 1 - \alpha - \int\displaylimits^{1}_{0} \int\displaylimits_{S_2(t_1) \cap [0, \infty)}  i\big(r\|\bs{b}\| k(t_{1}, \psi) ;\|\bs{b}\| \big) \, f_{R}(r) \,  dr \, dt_{1} \qquad \qquad \qquad \qquad \text{ for $q = 2$, } \\ 
1 - \alpha -\int\displaylimits^{1}_{0} \int\displaylimits^{1}_{0} \int\displaylimits_{S_q(t_1) \cap [0, \infty)} i \big(r\|\bs{b}\| k(t_{1}, t_{2}, \psi) ;\|\bs{b}\| \big) \, f_{R}(r) \,  dr \, f_{T_{1}}(t_{1}) \,  dt_{1} \, f_{T_{2}}(t_{2}) \, dt_{2}
\\
\qquad \qquad \quad
\qquad \qquad \qquad \qquad  
\qquad \qquad \qquad \qquad 
\qquad \qquad \qquad \qquad \quad \, \, \, \, \,\text { for $q > 2$}.
\end{dcases}
\end{align*}

Now, for each given $t_1$, $d(t_{1}, r; q, \|\blambda\|)$ is a quadratic function of $r$ with positive coefficient of $r^2$.
Let $r_{\rm min}(q, t_1)$ denote the value of $r$ that minimises $d(t_{1}, r; q, \|\blambda\|)$, for each given $t_1$. 
If $d \big(t_{1}, r_{\rm min}(q, t_1); q, \|\blambda\| \big) \ge \chi^{2}_{1-\widetilde{\alpha}, q}$
then $S_q(t_1) \cap [0, \infty)$ is either a single point or the empty set;
otherwise $S_q(t_1) = [l_q, u_q]$ where $l_q < r_{\rm min}(q, t_1) < u_q$.
It follows that if  
$d \big(t_{1}, r_{\rm min}(q, t_1); q, \|\blambda\| \big) < \chi^{2}_{1-\widetilde{\alpha}, q}$ then \newline
$[l_2, u_2] = \left[ -\|\blambda\| \cos(2 \pi t_{1}) \pm \left(\|\blambda\|^{2}\cos^{2}(2 \pi t_{1}) + \chi^{2}_{1-\widetilde{\alpha}, 2}  - \|\blambda\|^{2}\right)^{1/2}\right]$  and \newline
$[l_q, u_q] = \left[  -\|\blambda\| \cos( \pi t_{1}) \pm \left(\|\blambda\|^{2}\cos^{2}( \pi t_{1}) + \chi^{2}_{1-\widetilde{\alpha}, q}  - \|\blambda\|^{2}\right)^{1/2} \right]$
for $q >2$ .
Therefore
\begin{align*}
B_q
= 
\begin{dcases}
1 - \alpha -   \int\displaylimits^{1}_{0} \int\displaylimits_{[l_2, u_2]\cap [0, \infty)}  i \big(r \, \|\bs{b}\| \, k(t_{1}, \psi) ; \, \|\bs{b}\| \big) \, f_{R}(r)  \, dr \, dt_{1}   \quad \qquad \qquad \qquad \,    \, \,  \,  \text{ for $q = 2$, } \\ 
1 - \alpha -       \int\displaylimits^{1}_{0} \int\displaylimits^{1}_{0} \int\displaylimits_{[l_q, u_q]\cap [0, \infty)} i \big(r \, \|\bs{b}\| \, k(t_{1}, t_{2}, \psi) ; \, \|\bs{b}\| \big) \, f_{R}(r) \, dr \, f_{T_{1}}(t_{1}) \, dt_{1} \, f_{T_{2}}(t_{2})   \, dt_{2}   \\
\qquad \qquad \qquad \qquad   \qquad \qquad \qquad \qquad   \qquad \qquad \qquad \qquad  \qquad \qquad  \qquad \quad
\text { for $q > 2$}. 
\end{dcases}
\end{align*}
\hfill $\qed$

\medskip

\noindent{\it{Proof of Theorem 2}}

\medskip

Suppose that $\widetilde{\bgamma} = \bs{0}$ and that $1-\alpha$, $\widetilde{\alpha}$ and $\|\bs{b}\|$ are given. Also suppose that $\btheta$ is  given.
Thus $\blambda= \left(I^{\gamma \gamma}(\btheta,  \bs{0})\right)^{-1/2} \bgamma$.
We make the dependence of $\blambda$ on $\bgamma$
explicit with the notation $\blambda(\bgamma)$.
We also make the dependence of the probabilities $A$ and $B_q$, given by 
\eqref{eq:simplea} and \eqref{eq:bqjoint} respectively, on 
$\bgamma$
explicit with the notation $A(\bgamma)$ and $B_q(\bgamma)$.
Let $P_{\bgamma^{\prime}}(\, \cdot \, )$ denote a probability evaluated for the true parameter $\bgamma =\bgamma^{\prime}$. 

Using this notation, 
\begin{align*}
A(\bgamma)
& = P_{\bgamma}\left(\left| V_2 \right| \leq z_{1-\alpha/2}\right)  P_{\bgamma}\left(\|\bs{H}\|^2 \leq  \chi^{2}_{1-\widetilde{\alpha}, q} \right).
\end{align*}
Since $V_2 \sim N\left(- \bs{b}^T \blambda(\bgamma) /(1 - \|\bs{b}\|^2)^{1/2}, 1 \right)$, $P_{\bgamma^{\prime}}\left(\left| V_2 \right| \leq z_{1-\alpha/2}\right)  = P_{-\bgamma^{\prime}}\left(\left| V_2 \right| \leq z_{1-\alpha/2}\right)$. 
Also, note that
$\|\bs{H}\|^2$ has a noncentral $\chi^2$ distribution with $q$ degrees of freedom and noncentrality parameter $\|\blambda(\bgamma)\|^2$.
Since 
 $\|\blambda(\bgamma^{\prime})\|^2 = \|\blambda(-\bgamma^{\prime})\|^2$,  $A(\bgamma^{\prime}) = A(-\bgamma^{\prime})$.

Also
\begin{align*}
B_q(\bgamma) = P_{\bgamma} \big(| V_1 | \leq z_{1-\alpha/2},
\|\bs{H}\|^2 >  \chi^{2}_{1-\widetilde{\alpha}, q} \big).
\end{align*}
It follows from \eqref{eq:V1Hjointdist} that for true parameter value $\bgamma = \bgamma^{\prime}$,
\begin{align*}
\begin{bmatrix} 
-V_1 \\
-\bs{H}				
\end{bmatrix}
\sim
N
\left(
\begin{bmatrix}
0 \\
-\blambda(\bgamma^{\prime})
\end{bmatrix}
,
\begin{bmatrix}
1 & \bs{b}^T \\
\bs{b} & \bs{I}_q 
\end{bmatrix}
\right),
\end{align*}
and that for true parameter value $\bgamma = -\bgamma^{\prime}$,
$\big(V_1, \bs{H}^T \big)^T$ has the same distribution.
In other words, the distribution of $\big(V_1, \bs{H}^T \big)^T$ for the true parameter value $\bgamma = -\bgamma^{\prime}$ is the same as the distribution of 
$\big(-V_1, -\bs{H}^T \big)^T$ for the true parameter value $\bgamma = \bgamma^{\prime}$.
Therefore $B_q(\bgamma^{\prime}) = B_q(-\bgamma^{\prime})$.

\hfill $\qed$
}

\end{document}